\documentclass[letterpaper, 10 pt, conference]{ieeeconf}  

\IEEEoverridecommandlockouts                              
\usepackage{graphicx} 
\usepackage{float} 
\renewcommand{\vec}[1]{\mathbf{#1}}
\usepackage{amsmath} 
\usepackage{amsfonts} 
\usepackage{textcomp} 
\usepackage[caption=false]{subfig} 
\usepackage{subfig}
\usepackage{amssymb} 
\usepackage{multicol} 
\usepackage{cite}
\usepackage{bbm} 
\newtheorem{definition}{Definition}[section]

\title{\Huge 
Coverage and Rate Analysis of Super Wi-Fi Networks Using Stochastic Geometry
} 
\author{Neelakantan Nurani Krishnan, Gokul Sridharan, Ivan Seskar, Narayan Mandayam \\ WINLAB, Rutgers University\\ 671 US Route 1 South, North Brunswick, NJ - 08902 \\ \{\emph{neel45, gokul, seskar, narayan}\}@winlab.rutgers.edu} 
\begin{document}

\maketitle
\thispagestyle{empty}
\pagestyle{empty}

\begin{abstract}
\\
Recent regulatory changes proposed by the Federal Communications Commission (FCC) permitting unlicensed use of television white space (TVWS) channels present new opportunities for designing wireless networks that make efficient use of this spectrum. The favorable propagation characteristics of these channels and their widespread availability, especially in rural areas, make them well-suited for providing broadband services in sparsely populated regions where economic factors hinder deployment of such services on licensed spectrum. 
In this context, this paper explores the deployment of an outdoor Wi-Fi-like network operating in TVWS channels, referred to commonly as a Super Wi-Fi network. Since regulations governing unlicensed use of these channels allow (a) mounting fixed devices up to a height of 30 m and operation at transmit powers of up to 4 W EIRP, and (b) operation at transmit powers of up to 100 mW EIRP for portable devices, such networks can provide extended coverage and higher rates than traditional Wi-Fi networks. However, these gains are subject to the viability of the uplink from the portable devices (clients) to the fixed devices (access points (AP)) because of tighter restrictions on transmit power of clients compared to APs. This paper leverages concepts from stochastic geometry to study the performance of such networks with specific focus on the effect of (a) transmit power asymmetry between APs and clients and its impact on uplink viability and coverage, and (b) the interplay between height and transmit power of APs in determining the network throughput. Such an analysis reveals that (a) maximum coverage of no more than 700 m is obtained even when APs are deployed at 30 m height, and (b) operating APs at transmit power of more than 1 W is beneficial only at sparse deployment densities when rate is prioritized over coverage.    
\end{abstract}

\section{INTRODUCTION}
In light of rapidly growing mobile broadband traffic, providing additional spectrum is an important policy goal for spectrum regulators worldwide. With Internet of Things (IoT) and Machine to Machine (M2M) communications rising up the horizon, there is a need for ubiquitous connectivity. Meeting these requirements, especially in rural areas, is challenging because of geographic and monetary constraints. In this context, this paper investigates the viability of deploying an outdoor Wi-Fi-like network using television white space (TVWS) channels in rural/suburban areas for providing broadband connectivity. Such a network is typically referred to as a Super Wi-Fi network \cite{fcc}. 

\begin{figure}[t]
\centering
\includegraphics[width=3.35in]{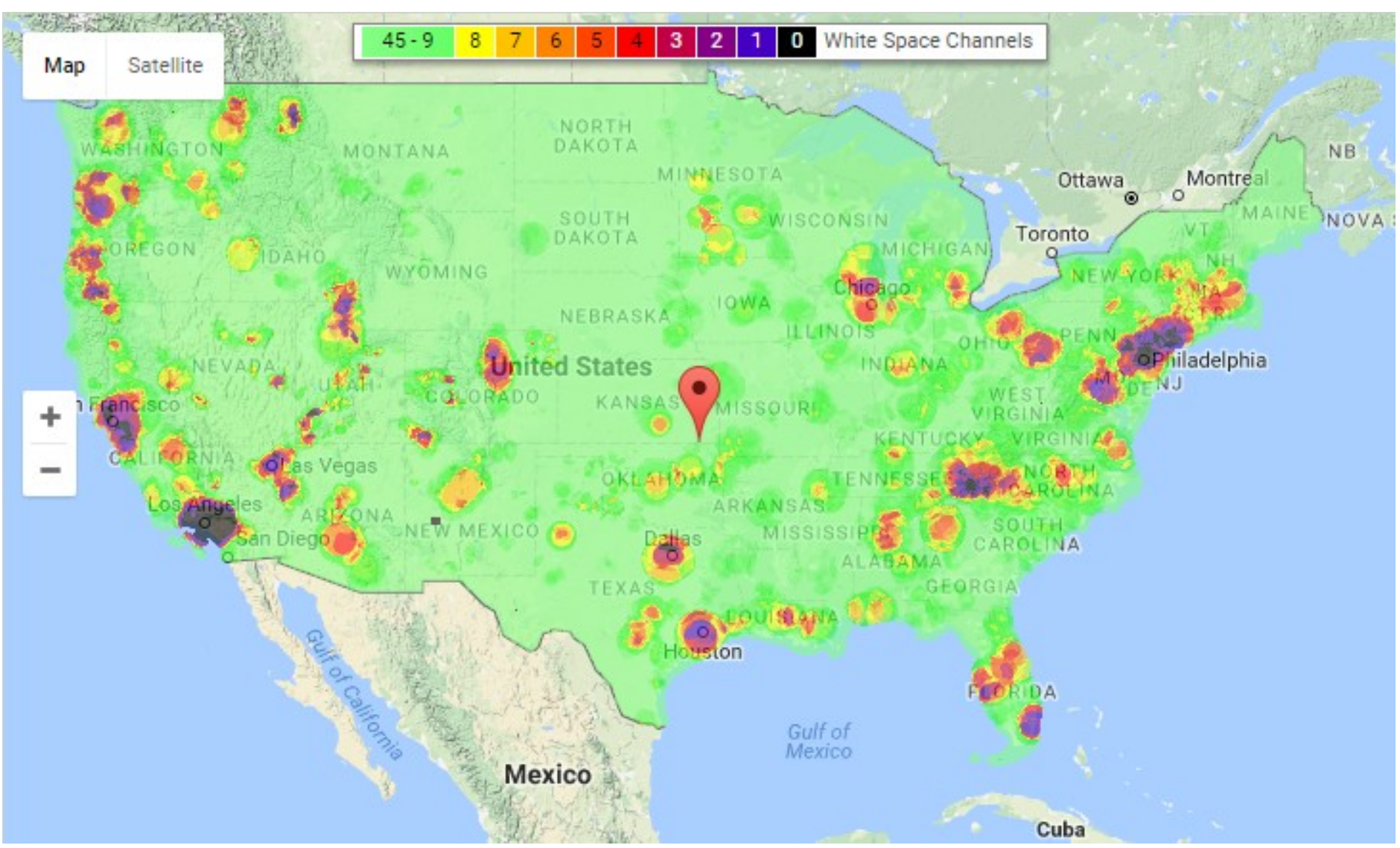}
\caption{Availability of TVWS Channels in the USA as of 02/15/2017 \cite{googlelink}. Except for the urban centers on either coasts, TVWS channels are abundantly available.}
\label{Spectrum Availability}
\end{figure}

TVWS channels are unused TV channels that can be opportunistically used on a secondary basis in the absence of primary transmissions. In the USA, these channels are 6 MHz wide and span from 54 MHz to 698 MHz. In particular, the channels in the 512-698 MHz range allow the secondary devices to be either fixed or portable, as in Table~\ref{allowed channels} \cite{fcc}. As seen in Fig.~\ref{Spectrum Availability}, TVWS channels are known to be locally under-utilized, especially in rural/suburban areas. Further, the relatively low frequency of TVWS channels comes with some significant advantages like lower path loss and better wall penetration \cite{11af}. 

\begin{table}[t]
\caption{TV channels for secondary use}
\label{allowed channels} 
\vspace{-0.4cm}
\begin{center}
\begin{tabular}{c c c c}
\hline \hline 
TV Channels & Freq. Band & Freq (MHz) & Allowed Devices\\
\hline
2 & VHF & 54-60 & Fixed\\
5-6 & VHF & 76-88 & Fixed \\
7-13 & VHF & 174-216 & Fixed \\
\hline 
14-20 & UHF & 470-512 & Fixed \\ 
21-35 & UHF & 512-602 & Fixed and Portable \\
39-51 & UHF & 620-698 & Fixed and Portable \\ 
\hline \hline 
\end{tabular}
\end{center}
\end{table} 
The under-utilization of TVWS channels in rural areas, along with their favorable propagation characteristics, motivates investigating the feasibility of deploying outdoor Super Wi-Fi networks  for broadband connectivity in such regions where providing access solutions continues to be exorbitantly expensive. In particular, this work envisions deploying a large number of fixed wireless access points (APs) over a rural/suburban area with channel access mediated by carrier sense multiple access with collision avoidance (CSMA/CA) while adhering to TVWS regulations. 
This work complements recent interest in utilizing TVWS channels for providing backhaul solutions \cite{cyrus,sudhir} where a wide-area network of cellular base stations over TVWS channels for backhaul is studied. 

To mitigate any potential impact on the primary users, secondary usage of TVWS channels is regulated by the Federal Communications Commission (FCC) in the USA. In particular, Table~\ref{regulations} specifies two important regulations that significantly impact the operation of a secondary Wi-Fi-like network---the first governs the maximum height of secondary devices and the second governs the maximum transmit power. In particular, the regulations allow mounting a fixed device at a height of up to 30 m while also allowing it to operate at up to 4 W EIRP (Equivalent Isotropically Radiated Power)---thus permitting a much larger coverage area when compared to a typical Wi-Fi AP that is mounted at much lower heights and is restricted to transmit at no more than 100 mW EIRP. However, since the portable devices are restricted to a much lower transmit power of 100 mW, an increase in downlink coverage from AP to client is not reciprocated by an equivalent increase in uplink coverage. In other words, there may be scenarios where the uplink may not be viable even though the downlink is. This presents a major point of difference between the network under consideration and traditional Wi-Fi networks. 
\begin{table}[t]
\caption{FCC regulations (maximum Values) for operation in TVWS channels}
\label{regulations} 
\vspace{-0.3cm}
\begin{center}
\begin{tabular}{|c| c|}
\hline
Fixed device height & 30 m\\
\hline
Portable device height & 1.5 m\\
\hline 
Fixed device EIRP & 4 W (per channel) \\
\hline 
Portable device EIRP & 100 mW (per channel) \\
\hline 
\end{tabular}
\end{center}
\end{table} 

Given the crucial role played by uplink association request and acknowledgment packets in determining AP-client association and successful downlink transmissions, the significant difference in operating parameters between APs and clients makes it extremely important to factor in uplink viability. Additionally, the potentially large downlink coverage can have a detrimental impact on the AP transmission probabilities when using CSMA/CA. Thus, although an increase in AP transmit power and/or height might seem beneficial, the above issues highlight the difficulty in choosing the right set of operating parameters for striking the right balance between coverage and throughput in such networks. 

With the broad goal of understanding the performance of large-scale outdoor Super Wi-Fi networks, this work uses concepts from stochastic geometry to obtain an analytical characterization of SINR coverage and rates. Tools from stochastic geometry are used to first characterize the probability of transmission of a Super Wi-Fi AP and subsequently study the area spectral efficiency (ASE) of such a network as a function of (a) AP deployment density, (b) AP height, and (c) AP transmission power. Due to the unequal transmission powers between APs and clients, this analysis explicitly requires the uplink to be viable when computing the downlink throughput. It is primarily in this respect that the current work significantly differs from existing literature on analyzing such networks \cite{main_paper,new_insights,notonsteroids}. 


The results of this work show that it is not always beneficial to operate at high AP transmit power ($P_{AP}$) and AP height ($h_{AP}$). It is consistently observed across different deployment densities that operating at high $P_{AP}$ and $h_{AP}$ values leads to a sharp drop in probability of transmission for APs, in turn decreasing the ASE. At deployment densities of less than $1 \,\text{AP}/\mathrm{km^2}$, ASE is maximized when $P_{AP}$ and $h_{AP}$ are close to 1 W and 1.5 m, respectively. At higher deployment densities of $10\, \text{APs}/\mathrm{km^2}$, ASE is maximized when $P_{AP}$ and $h_{AP}$ are close to 0.1 W and 1.5 m, respectively. On the contrary, for optimal coverage, maximizing $h_{AP}$ proves beneficial (note that coverage is determined by uplink viability and hence is independent of $P_{AP}$). Setting $P_{AP}$ greater than 1 W is observed to be useful only in sparse deployment densities, when rate is prioritized over coverage, with APs mounted at less than 6 m height. 

The rest of the paper is organized as follows --- related work is presented in Section~\ref{related_work}, the system model and parameters involved in the analysis are described in Section~\ref{system_model}, characterization of the network throughput is given in Section~\ref{throughput_analysis}, a discussion of the results obtained is provided in Section IV and the concluding remarks are mentioned in Section~\ref{conclusion}.  

\section{RELATED WORK} 
\label{related_work}
The release of TVWS channels for unlicensed use promoted active research in investigating the feasibility of deploying cognitive radio networks in these channels \cite{technical_challenges,Sahai,quantify,Europe}. The possibility of using TVWS channels for Super Wi-Fi operation is explored extensively in \cite{stirling,whitefiarchi,whitefi2,notonsteroids,Africa}. The authors of \cite{whitefiarchi} propose to enhance the coverage of public Wi-Fi networks operating in 2.4 GHz by extending their operation to TVWS channels. The authors of \cite{whitefi2} built a prototype for Super Wi-Fi (called White-Fi) networking and modified the medium access control (MAC) protocol to factor in spatial and temporal variations of TVWS channels. In \cite{notonsteroids}, a quantitative study of Super Wi-Fi networks is provided and it is observed that TVWS is an attractive alternative for providing connectivity in outdoor rural areas. A real world deployment of Super Wi-Fi networks is presented in \cite{Africa} where the potential of using TVWS for bringing broadband connectivity to unconnected areas is established. The issue of transmit power asymmetry in TVWS networks has been studied in \cite{scout}, albeit in a vehicular connectivity set up, in which the authors propose to extend the range of uplink from clients by using existing cellular paths. However, in the current work, rural areas are the target environments and hence no form of connectivity is likely to be preexistent. While these efforts provide the motivation to better utilize this technically and economically significant frequency range, a theoretical study of the performance of a large scale Super Wi-Fi network is not available to the best of our knowledge. Additionally, such an analysis should consider the various regulatory constraints imposed on the operating parameters, specifically transmission powers and heights, as described in Table~\ref{regulations}. The current work provides a theoretical framework to serve this purpose by employing concepts from stochastic geometry. 



Among the first efforts to theoretically analyze traditional Wi-Fi networks were the analyses presented in \cite{wlan} and \cite{bianchi} to accurately model the 802.11 protocol. While these efforts captured finer aspects of the CSMA/CA protocol (e.g., exponential backoff), spatial aspects of the wireless medium are not modeled. Stochastic geometry provides a natural framework to analyze wireless networks while retaining the spatial characteristics of signal propagation. The use of stochastic geometry for modeling and analyzing wireless networks started with the extensive analysis of ALOHA \cite{aloha1,aloha2}. In particular, \cite{aloha2} studies CSMA-based networks using a Matern hard-core point processes where each AP was assumed to have a disc of fixed radius around itself within which there are no other APs. A modification to this analysis that modeled the backoff procedure in CSMA/CA and included fading was presented in \cite{main_paper,csma1,new_insights}. The basic framework of \cite{main_paper} to analyze CSMA/CA forms the foundation for the current effort. Subsequent analysis of interference due to concurrent AP transmissions is modeled using the methodology proposed in \cite{ganti,andrews}. 





A comprehensive overview of using stochastic geometry to model a wide variety of wireless networks is given in \cite{seminal_stoc,stoc1}. The mathematical tools and theory of point processes used in the current analysis are presented in \cite{text1} and \cite{text2}. 
\section{SYSTEM MODEL} 
\label{system_model}
Consider a large set of APs whose locations are fixed and drawn from a homogeneous Poisson point process (PPP) of intensity $\lambda$. The set of AP locations is given by $\Phi_A$ = $\left\{\vec{x}_1, \vec{x}_2, ..., \vec{x}_k, ...\right\}.$ The notation $||(\vec{x}_i-\vec{x}_j)||$ is used to represent distances between APs at two locations $\vec{x}_i$ and $\vec{x}_j$. The APs are assumed to only serve clients located within its Voronoi cell (provided the uplink from client to AP is viable), with client locations being uniformly distributed within the Voronoi cell. In such a setting, the distribution of AP-Client distance $\mathnormal{r}$ (without factoring uplink viability) is given by 
\begin{equation} 
\mathnormal{f}_r(\mathnormal{r}) =2\pi \lambda\mathnormal{r}e^{-\lambda\pi \mathnormal{r}^2}.
\end{equation}
It is assumed that all APs have at least one associated client to serve at any instance. The APs have access to one TVWS channel (6 MHz wide) and all APs contend to get access to this channel. APs are bound by TVWS regulations that govern fixed devices while the clients are bound by the regulations governing portable devices. Thus, while APs can transmit at any transmit power $P_{AP} \leq  4 \,\mathrm{W}$, the clients are assumed to transmit at $P_{C} = 0.1\,$W. The APs are also assumed to be mounted at any height $h_{AP}\leq 30 \, \mathrm{m}$, while clients are always assumed to be at $1$ m height. Isotropic antennas are assumed at both APs and clients. The analysis in this paper assumes a persistent downlink traffic and negligible uplink traffic.  
\subsection{Radio Propagation Model} 
The power received at a point $\vec{y}$ from an AP located at $\vec{x}$ is given by  
\begin{equation} 
P(\vec{x},\vec{y}) = P_{AP}\,\rho(\vec{x},\vec{y})\,F(\vec{x},\vec{y}),
\end{equation} 
where $\rho(\vec{x},\vec{y})$ is the pathloss encountered by the transmission between $\vec{x}$ and $\vec{y}$, and $F(\vec{x},\vec{y})$ is the fading coefficient between $\vec{x}$ and $\vec{y}$. $F(\vec{x},\vec{y})$ is modeled as an i.i.d. exponential random variable with mean $\mu = 1$. The notation P($d$), $\rho(d)$ and F($d$) are used when referring to received power, pathloss and fading coefficient between two generic locations that are at a distance `$d$' from each other. 

Two different pathloss models are used to define AP-AP and AP-Client transmission links. The two pathloss models are drawn from the dual-slope model specified in \cite{pathloss} for suburban environments. This particular pathloss model is chosen as it is sensitive to transmitter and receiver heights and is applicable to a wide range of sub-GHz frequencies. The dual-slope model is given by 
\begin{equation} 
\rho(\vec{x},\vec{y}) \, \text{in $\mathrm{dB}$} = 
\begin{cases} 
\! \rho_{\text{LOS}} + 20 + 25 \log\left(\tfrac{d}{R_{bp}}\right)\!,\, \text{if}\ d < R_{bp} \\ 
\! \rho_{\text{LOS}} + 20 + 40 \log\left(\tfrac{d}{R_{bp}}\right)\!,\, \text{if}\ d \geq R_{bp} 
\end{cases}
\end{equation}
where 
\begin{itemize}
\item $d$ is the distance between $\vec{x}$ and $\vec{y}$,
\item $\rho_{\text{LOS}}$ is the line-of-sight pathloss (in dB), given by

\begin{equation} 
\rho_{\text{LOS}} = \left| 20 \log\left(\frac{\lambda^2}{8\pi h_t h_r} \right)\right|,
\end{equation} 
\item $R_{bp}$ is the breakpoint distance (in meters), given by
\begin{equation} 
R_{bp} = \frac{1}{\lambda}\sqrt{(\Sigma^2 - \Delta^2)^2 - 2(\Sigma^2 + \Delta^2)\left(\frac{\lambda}{2}\right)^2 + \left(\frac{\lambda}{2}\right)^4}.
\end{equation}
\end{itemize}
The different parameters involved in the above model are
\begin{itemize} 
\item $\lambda$ - wavelength (m),  
\item $h_t$ - height of transmitting antenna (m), 
\item $h_r$ - height of receiving antenna (m),
\item $\Sigma = h_t + h_r$,
\item $\Delta = h_t - h_r$.
\end{itemize}
For modeling AP-AP transmissions, $h_t$ and $h_r$ are set to AP antenna height $h_{AP}$. For modeling AP-client transmissions, $h_t = h_{AP}$ and $h_r = 1\, \mathrm{m}$. 


\subsection{Channel Contention Model} 
Channel access in the current network is governed by CSMA/CA. In CSMA/CA, an AP gets access to a channel when there are no other contending APs in its neighborhood (i.e., the channel is sensed to be idle), otherwise an exponential back-off procedure is initiated. The channel is sensed to be idle when the received signal strength from all neighboring APs is below the clear-channel-assessment (CCA) threshold. In conventional Wi-Fi networks, the CCA threshold is typically set to -82 dBm and the same threshold is used in the current work.  

\subsection{Uplink Viability} 
Uplink viability determines the ability of a client to associate with a neighboring AP. If the association request messages from the client do not reach an AP, the client cannot be served. This scenario may often occur in Super Wi-Fi networks as clients transmit at powers lower than APs. Note that even though uplink viability also affects the receipt of acknowledgment (ACK) packets from the client indicating successful downlink transmission, the assumption in the current work is that once association is established between an AP-client pair, the channel remains time invariant. This paper defines uplink viability as follows.
\begin{definition}
The uplink transmission between client $\vec{y}$ and its AP $\vec{x}$ is said to be viable if the received signal strength from the client to the AP exceeds a certain threshold $\gamma$, i.e.,
\begin{equation}
P_{C}\,\rho(\vec{x},\vec{y})\,G(\vec{x},\vec{y}) > \gamma.
\label{uplink_def}
\end{equation}
\end{definition} 

Note that channel reciprocity is not assumed and hence $G(\vec{x},\vec{y})$ and $F(\vec{x},\vec{y})$ are two independent random variables. In this paper, the threshold $\gamma$ is set to be equal to the CCA threshold $\sigma$. Although the criterion (\ref{uplink_def}) only accounts for uplink packet detection and not successful decoding, it simplifies the subsequent analysis while retaining the essential characteristics of the network under consideration.

Using the above definition, uplink viability is computed as 
\begin{equation} 
p_{U}(r) = \mathbb{P}[P_{C}\, \rho(\vec{x},\vec{y})\, G(\vec{x},\vec{y}) > \gamma] = e^{-\frac{\mu\gamma}{P_{C}\rho(\vec{x},\vec{y})}}.
\label{pu_def}
\end{equation} 
where $r = ||\vec{x} - \vec{y}||$ is the distance between AP $\vec{x}$ and client $\vec{y}$. 

In particular, coverage range of an AP is defined as the largest AP-client distance d such that 
\begin{equation} 
p_U(d) \geq 0.1  
\end{equation}
where $p_U(r)$ is defined in (\ref{pu_def}).

\subsection{Transmission Model} 
The signal to interference-plus-noise ratio (SINR) observed at a client $\vec{y}$ and associated with an AP $\vec{x}$ is given by 
\begin{equation} 
\mathrm{SINR}(\vec{x},\vec{y}) = \frac{\rho(\vec{x},\vec{y})}{N_0 + \sum_{\vec{z}\in\Phi_T\setminus{\vec{x}}}I(\vec{z},\vec{y})}
\end{equation}
where
\begin{itemize} 
\item $N_0$ is the noise variance,
\item $\Phi_T$ is the set of concurrently transmitting APs,
\item $\sum_{\vec{z}\in\Phi_T\setminus{\vec{x}}}I(\vec{z},\vec{y})$ is the cumulative interference at client $\vec{y}$ due to all concurrently transmitting APs except AP $\vec{x}$. 
\end{itemize}
The transmitted data rate from AP $\vec{x}$ to client $\vec{y}$ is then given by $\log_2(1+\text{SINR}(\vec{x},\vec{y}))$.

\section{THROUGHPUT MODELING AND ANALYSIS}  
\label{throughput_analysis}
This section focuses on characterizing the performance of the network under consideration through metrics such as AP transmission probability and area spectral efficiency. Area spectral efficiency is defined as the average throughput of an AP multiplied by the density of the AP deployment. It is assumed that all APs actively contend for the channel and when channel access is granted, use the channel for a fixed period of time to transmit to one of their associated clients. For an AP to serve a client the following three transmissions must be successfully received: (a) association request packets at the AP, (b) transmission payload packets at the client, and (c) acknowledgment packets at the AP. Clearly, (a) and (c)  both require uplink viability, and it is assumed that as long as the uplink received signal strength exceeds the threshold $\gamma$, both these transmissions are successful. Under these assumptions, the probability of an AP serving a user at a distance $r$ is equal to the probability of an uplink viable client being located at a distance r and is given by
\begin{equation}
f_R(r|I_u=1)=\frac{f_R(r)\mathbb{P}{(I_u=1|R=r})}{\mathbb{P}{(I_u=1)}}, 
\label{equation_fr}
\end{equation}
where $I_u$ is a binary random variable indicating uplink viability. In particular, $I_u = \mathbbm{1}(P_{C}\,\rho(r)\,G(r) > \gamma)$, and

\begin{equation}
\mathbb{P}{(I_u=1|R=r})=\mathbb{P}\left (G(r) > \tfrac{\gamma}{P_{C}\,\rho(r)} \right )=e^{\frac{-\mu\gamma}{P_c\, \rho(r)}}.
\end{equation}

Given the random deployment of APs and clients, the average throughput is computed over all possible AP and and client locations. In particular, when the AP has channel access and serves an uplink-viable client located at a distance $r$, the average throughput to that client is given by

\begin{equation}
\mathcal{T}(r)=\mathbb{E}_{(\text{SINR}|R=r)}[\log(1+\mathrm{SINR})] 
\label{tpt-1}
\end{equation}
where the expectation is over the distribution of SINR conditioned on the client being at a distance of $r$ from the AP. Note that the SINR distribution is independent of uplink viability.

Thus, the average throughput of an AP after accounting for the probability of transmission can be written as
\begin{equation} 
\mathcal{T} = \int_{0}^{\infty} \mathnormal{p_T(r)} \mathcal{T}(r) f_R(r|I_u=1)  d\mathnormal{r} 
\label{tpt-2}
\end{equation}
where $p_T(\mathnormal{r})$ is the transmission probability of the AP conditioned on serving a client that is at a distance of $\mathnormal{r}$. Just as the SINR distribution, $p_T(\mathnormal{r})$ is also independent of uplink viability.

The rest of the section focuses on computing $p_T(\mathnormal{r})$ and $\mathcal{T}(r)$. The methodology adopted is similar to the framework presented in \cite{main_paper,new_insights}.

\subsection{Probability of AP Transmission: $p_T(r)$} 
Probability of an AP transmitting is governed by CSMA/CA and the exponential backoff procedure. As proposed in \cite{main_paper}, the exponential backoff procedure used by an AP when the channel is busy can be approximately modeled by tagging each AP in the Poisson field with an independent mark. This mark decides the backoff time of that AP. 

In particular, each AP $\vec{x}$ in $\Phi_A$ is assigned an independent mark $m_{\vec{x}}$ uniformly distributed in [0,1]. Defining the neighborhood of an AP $\vec{x}$ as $\mathcal{N}(\vec{x}) = \left\{\vec{y} \in \Phi_A : P(\vec{x},\vec{y}) > \sigma\right\}$, an AP transmits if no other AP in its neighborhood has a smaller mark than itself. Thus, the set of concurrently transmitting APs can now be defined as
\begin{equation} 
\Phi_\mathnormal{T} = \left\{\vec{x} \in \Phi_A : \mathnormal{m}_\vec{x} < \mathnormal{m}_\vec{y}, \forall \, \vec{y} \in \mathcal{N}(\vec{x}) \right\}.
\end{equation} 

This model captures the fact that CSMA/CA grants channel access to that AP with minimal back-off time (equivalent to having lowest mark) among all APs in its neighborhood and that an AP abstains from transmitting if another AP in its neighborhood is already transmitting.

Note that this approximate model ignores collisions, the exponential nature of back-off, and the history of timers. Nevertheless, as shown by the authors in \cite{main_paper}, through ns-2 simulations, this model provides fairly accurate results. 

Without loss of generality, we focus on an AP located at the origin and denoted as AP $\vec{0}$. Let AP $\vec{0}$ serve a client $\vec{y}$ located at a distance  $r$. Computing the transmission probability is equivalent to computing the probability that among the APs with a mark less that $m_{\vec{0}}$, none of them are in the neighborhood, i.e.,
\begin{equation}
p_T(\mathnormal{r})=\mathbb{P}\left (m_\vec{x}\geq m_\vec{0} \ \forall \ \vec{x} \in \mathcal{N}(\vec{0}) \right ).
\end{equation} 

It can be shown using the results in  \cite{main_paper,new_insights} that the transmission probability, as defined above, can be computed as
\begin{align}
\mathnormal{p_T(r)} =& \int_{0}^{1} e^{-\lambda m_0\int_{\mathbb{R}^2\setminus\mathcal{B}(\vec{y},\mathnormal{r})} \mathnormal{S}(\vec{x}) d\vec{x}} dm_0 \label{pt_eq1} \\
=& \frac{1-e^{-\lambda\int_{\mathbb{R}^2\setminus\mathcal{B}(\vec{y},\mathnormal{r})}\mathnormal{S}(\vec{x})d\vec{x}}}{\lambda\int_{\mathbb{R}^2\setminus\mathcal{B}(\vec{y},\mathnormal{r})}\mathnormal{S}(\vec{x})d\vec{x}}
\label{pt_eq2}
\end{align}
where $\mathcal{B}(\vec{y},\mathnormal{r})$ is a ball of radius $\mathnormal{r}$ with the client at its center (which by hypothesis cannot contain any AP other than AP $\vec{0}$) and $\mathnormal{S}(\vec{x})$ is the probability of AP $\vec{0}$ detecting an AP at $\vec{x}$. $\mathnormal{S}(\vec{x})$ can be computed as
\begin{align} 
\mathnormal{S}(\vec{x}) = \mathbb{P}[\mathnormal{P_{AP}\,\rho}(\vec{0},\vec{x})\,\mathnormal{F}(\vec{0},\vec{x}) > \sigma] = e^{\frac{-\mu \sigma}{P_{AP}\,\rho(\mathnormal{||x||})}}
\end{align}

The expression in (\ref{pt_eq2}) can be computed in a straightforward manner using standard numerical techniques.

\subsection{Average Throughput to a Client: $ \mathcal{T}(r) $}
Computing the average throughput delivered by an AP to its associated client at a distance $r$ requires determining the distribution of SINR at the client. This in turn requires the computation of the cumulative interference caused at the client due to all other APs concurrently transmitting with AP $\vec{0}$. To compute the SINR distribution, the methodology used in \cite{new_insights} is adopted.

In particular, the complementary cumulative distribution function (CCDF) of SINR can be expressed using Laplace functionals and written as
\begin{equation}
\mathbb{P}(\mathrm{SINR}(r)> \beta) = \psi_I(\mathnormal{s})\psi_N(\mathnormal{s}),
\label{laplace_eq}
\end{equation}
where $s = \frac{\mu\beta}{P_{AP}\rho(\mathnormal{r})}$, and $\psi_I(\cdot)$ and $\psi_N(\cdot)$ are the Laplace functionals of the interference from other AP transmissions and additive noise, respectively.

Switching to a user-centric perspective by shifting the origin to the location of client and assuming AP $\vec{0}$ to now be located at $(r,0)$ (in polar coordinates), (\ref{laplace_eq}) can be approximated as
\begin{equation} 
\mathbb{P}(\mathrm{SINR}> \beta)\hspace{-0.3mm} \approx \hspace{-0.3mm} e^{\mathnormal{-sN_0}}e^{-\lambda\int_{0}^{2\pi}\hspace{-0.5mm}\int_{\mathnormal{r}}^{\infty}\mathnormal{q(b(v,\theta))[1-\phi_F(s\rho(v))]v dv d\theta}} 
\label{approx_eq}
\end{equation} 
where $s$ is as before, and
\begin{itemize}  
\item $q(d)$ is the probability that two APs separated by a distance $d$ transmit concurrently (computation of $q(d)$ is given in Appendix I),
\item $b(v,\theta) = v^2 + \mathnormal{r}^2 - \mathnormal{r}v cos(\theta)$ is the distance between the serving AP at $(r,0)$ and a generic interfering AP located at $(v,\theta)$, 
\item $\phi_\mathnormal{F}$ - Laplace transform of the fading random variable \\ $\phi_\mathnormal{F}(\mathnormal{x}) = \frac{1}{1+\mathnormal{x}}$ as fading is exponentially distributed, 
\item $N_0$ - Thermal noise variance. 
\end{itemize} 

Evaluating (\ref{approx_eq}) at $s = \frac{\mu\beta}{P_{AP}\rho(\mathnormal{r})}$ using standard numerical techniques yields the distribution of SINR at the client. Once the distribution of SINR is obtained, the expected rate delivered by the AP can be computed as $\mathbb{E}[\log(1+\text{SINR})])$ where the expectation is computed over the distribution of SINR. 

\section{NUMERICAL RESULTS AND DISCUSSION} 
\label{results}
This section presents the results obtained using the methodology outlined earlier and highlights key takeaways on the design of Super Wi-Fi networks. Some of the key parameters used in the computations are given in Table~\ref{setup}. 
\begin{table}[H]
\caption{Deployment set up}
\label{setup} 
\vspace{-0.5cm}
\begin{center}
\begin{tabular}{c|c }
\hline \hline
Channel Center Frequency & 600 MHz \\
Channel Bandwidth & 6 MHz \\ 
Number of channels & 1 \\
AP distribution & Homogeneous PPP of intensity $\lambda$ \\  
AP Transmission Power & Variable between 0.1 W and 4 W\\
Client Transmission Power & 0.1 W \\ 
AP Height & Variable between 1.5 m and 30 m \\
Client Height & 1 m \\
Pathloss & Based on ITU-R P.1411-8\\
Fading & Exponentially distributed with mean 1 \\
CCA Threshold $\sigma$ & -82 dBm \\ 
Uplink Viability Threshold $\gamma$ & -82 dBm \\ 
Noise Variance ($N_0$) & -173.97 $\mathrm{dBm/Hz}$ \\
Traffic model & Persistent downlink \\
\hline \hline 
\end{tabular}
\end{center}
\label{params_table}
\end{table} 
\subsection{Validation of Methodology} 
To validate the proposed stochastic-geometry-based model of Super Wi-Fi networks, results obtained using such an approach are compared against simulation results obtained using OPNET, an industry-standard packet-based network simulation tool \cite{opnet}.

In particular, due to computational complexity of large scale simulations in OPNET and lack of in-built support for rate adaptivity, OPNET is used to simulate the performance of a network with a single AP and a single client. This result is then compared against results obtained using the stochastic-geometry-based model in a sparse deployment setup where effects of channel contention are minimized and the AP transmissions can be assumed to be independent of each other. 

For the OPNET simulation $h_{AP}$ and $P_{AP}$ are set to 30 m and $0.1$ W or 1 W, respectively. Due to lack of support for rate adaptivity, modulation-and-coding-scheme (MCS) index was varied manually to identify the best index for a given setup. While the pathloss model used in OPNET is different from the one listed Table III, the simulation tool is used as a means of validating the stochastic geometry analysis developed in this paper. Specifically, OPNET uses the Suburban Hata model, where the pathloss after substitution of $h_{AP}$ = 30 m is given by $\rho(d)= 124.3 + 35.23\log(d)$, where $d$ is in km. For comparison, the stochastic-geometry-based model is also set to use the Suburban Hata model with AP density set to 0.1 AP/km$^2$. 
\begin{figure}[t]
\centering 
\subfloat[AP throughput vs. AP-client distance $\mathnormal{r}$ for $P_{AP} = 0.1\, \mathrm{W}$ and $h_{AP} = 30\, \mathrm{m}$]{\label{OPNET_vs_SG_0_1}\includegraphics[scale = 0.4]{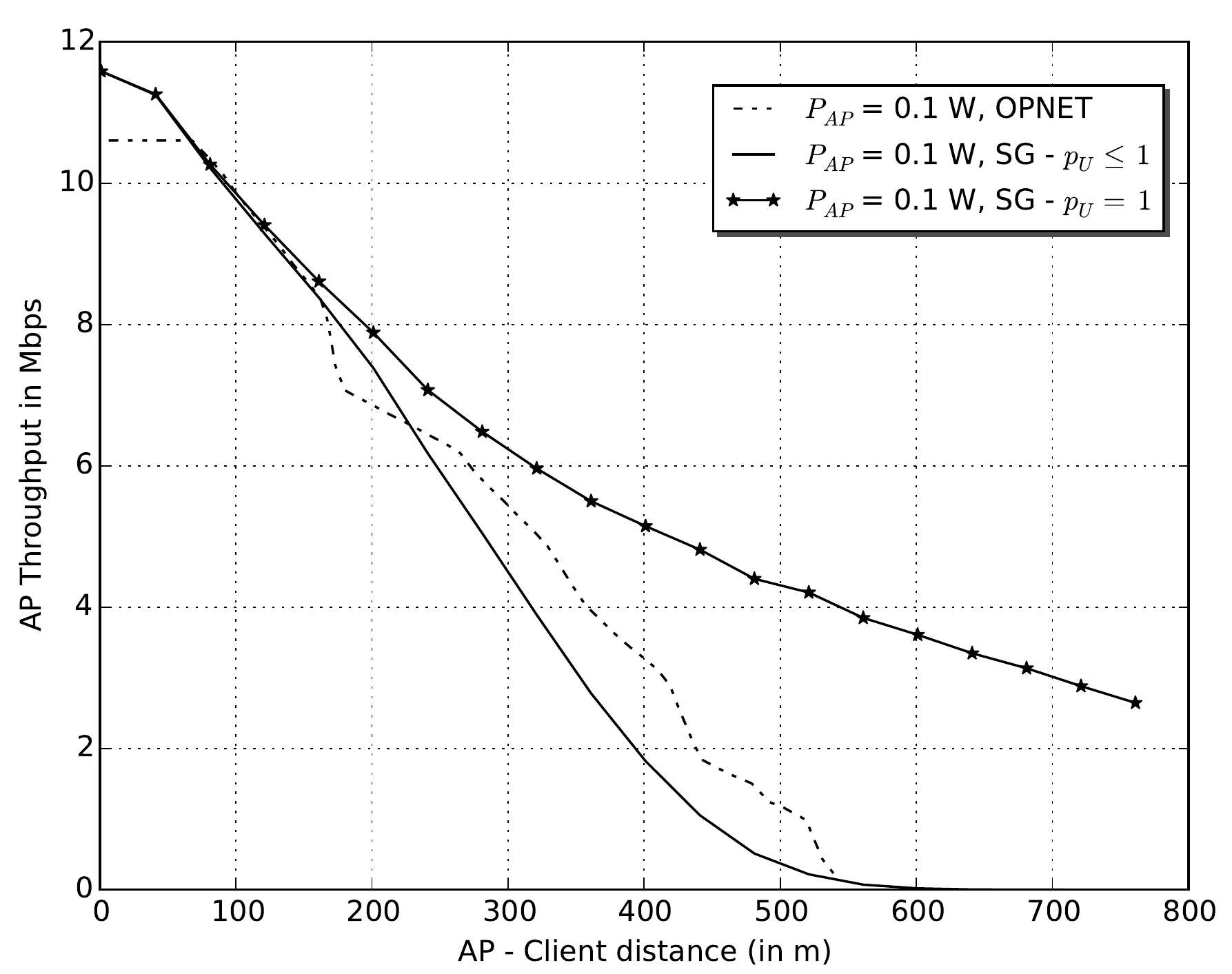}}  \newline
\subfloat[AP throughput vs. AP-client distance $\mathnormal{r}$ for $P_{AP} = 1\, \mathrm{W}$ and $h_{AP} = 30\, \mathrm{m}$]{\label{OPNET_vs_SG_1}\includegraphics[scale = 0.4]{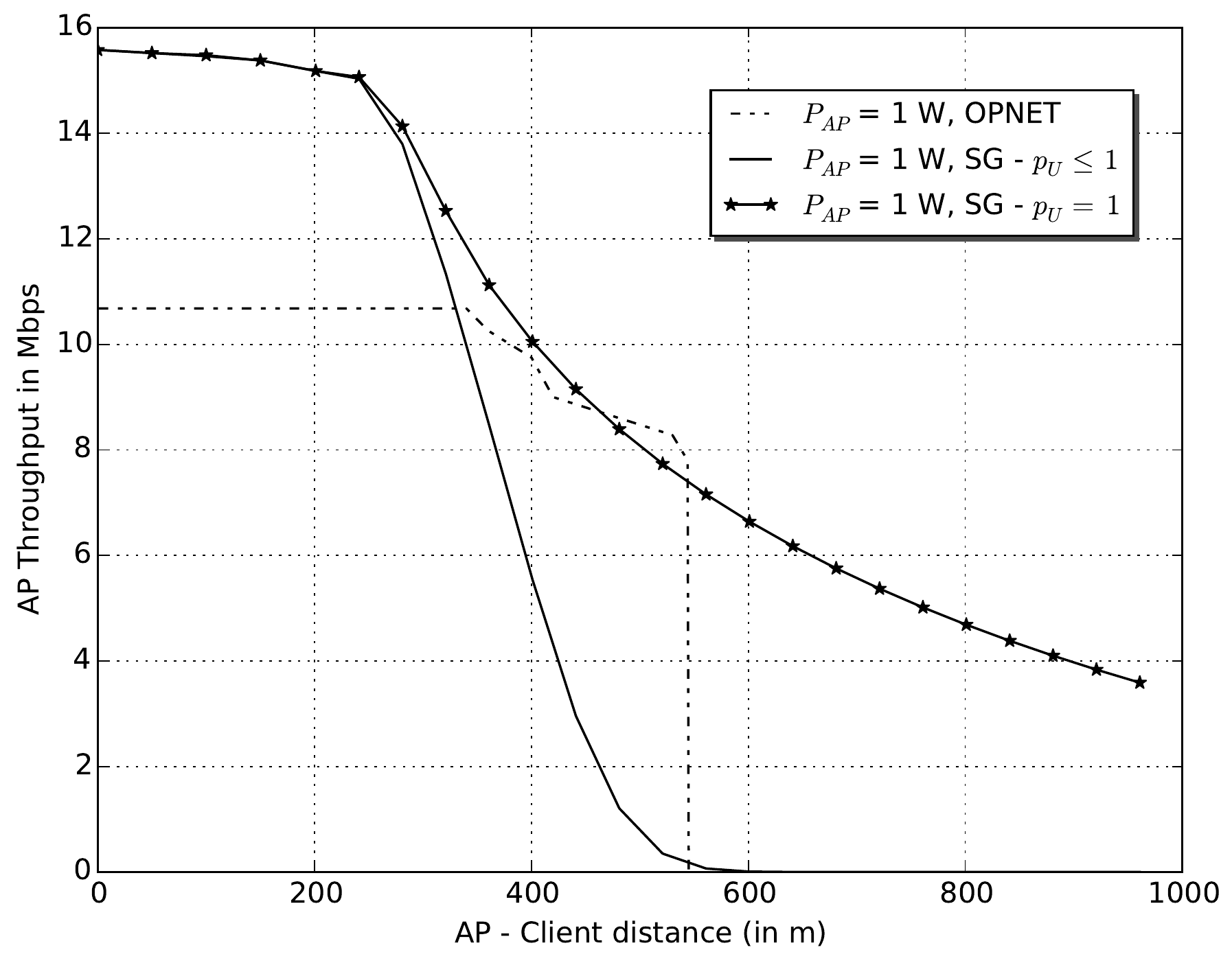}}
\caption{Results from stochastic-geometry-based models and OPNET simulation. The plot with $p_U \leq 1$ is for the case when uplink viability is probabilistic and the one with $p_U = 1$ is when uplink is assumed to be always viable.}
\label{OPNET_vs_SG}
\end{figure} 


Fig.~\ref{OPNET_vs_SG} shows the plots of AP throughput (in Mbps) per channel as a function of AP-client distance (in m) for three cases - (i) OPNET simulation, (ii) stochastic-geometry model when uplink viability is factored in, and (iii) stochastic-geometry model under the assumption that uplink is always viable. Note that in OPNET fading is not modeled, which is evident from the abrupt fall in throughput at $\sim 550$ m in Fig.~\ref{OPNET_vs_SG_1}. It is seen from Fig.~\ref{OPNET_vs_SG_0_1} that the current model closely follows the throughput obtained using OPNET. Further, even though uplink and downlink powers are the same in this case, factoring uplink viability produces more accurate results when compared to the existing models for Wi-Fi networks. The impact of uplink viability is even more pronounced in Fig.~\ref{OPNET_vs_SG_1}. Note that the proposed model and the OPNET results both indicate that clients beyond $550$ m are incapable of being served. At shorter distances, while restrictions on MCS indices cap the maximum throughput in OPNET, no such restriction is placed on the proposed model. Due to lack of fading in OPNET simulations, OPNET predicts larger throughputs in the 400 m to 550 m range than the proposed model. It is clear that models that do not factor in uplink viability are particularly inaccurate at larger distances and predict much larger coverage than is realistically possible. These results along with the comparison in \cite{main_paper} against ns-2 simulations further validate this model.
\subsection{Results on Throughput and Coverage Analysis} 
The section presents the projected performance of Super Wi-Fi networks from the perspective of (a) transmission probability, (b) coverage of an AP, and (c) area spectral efficiency (network throughput). A well-designed Super Wi-Fi network requires striking the right balance between all of the above three attributes.

\begin{figure*}[t] 
\centering 
  \subfloat[Low Density]{\label{pt_0_1}\includegraphics[width=5.9cm,height=6cm]{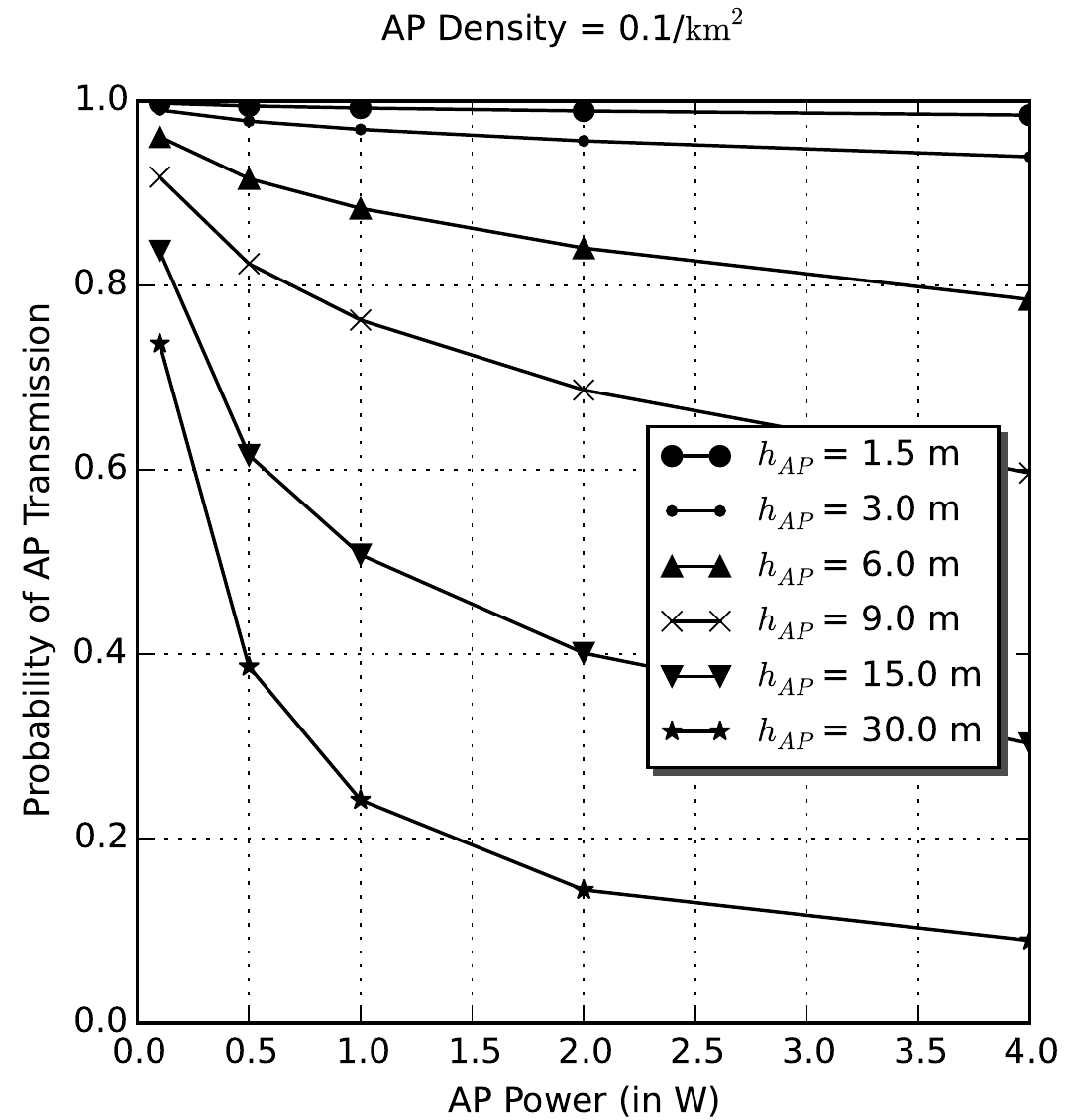}} \hfill
  \subfloat[Medium Density]{\label{pt_1}\includegraphics[width=5.9cm,height=6cm]{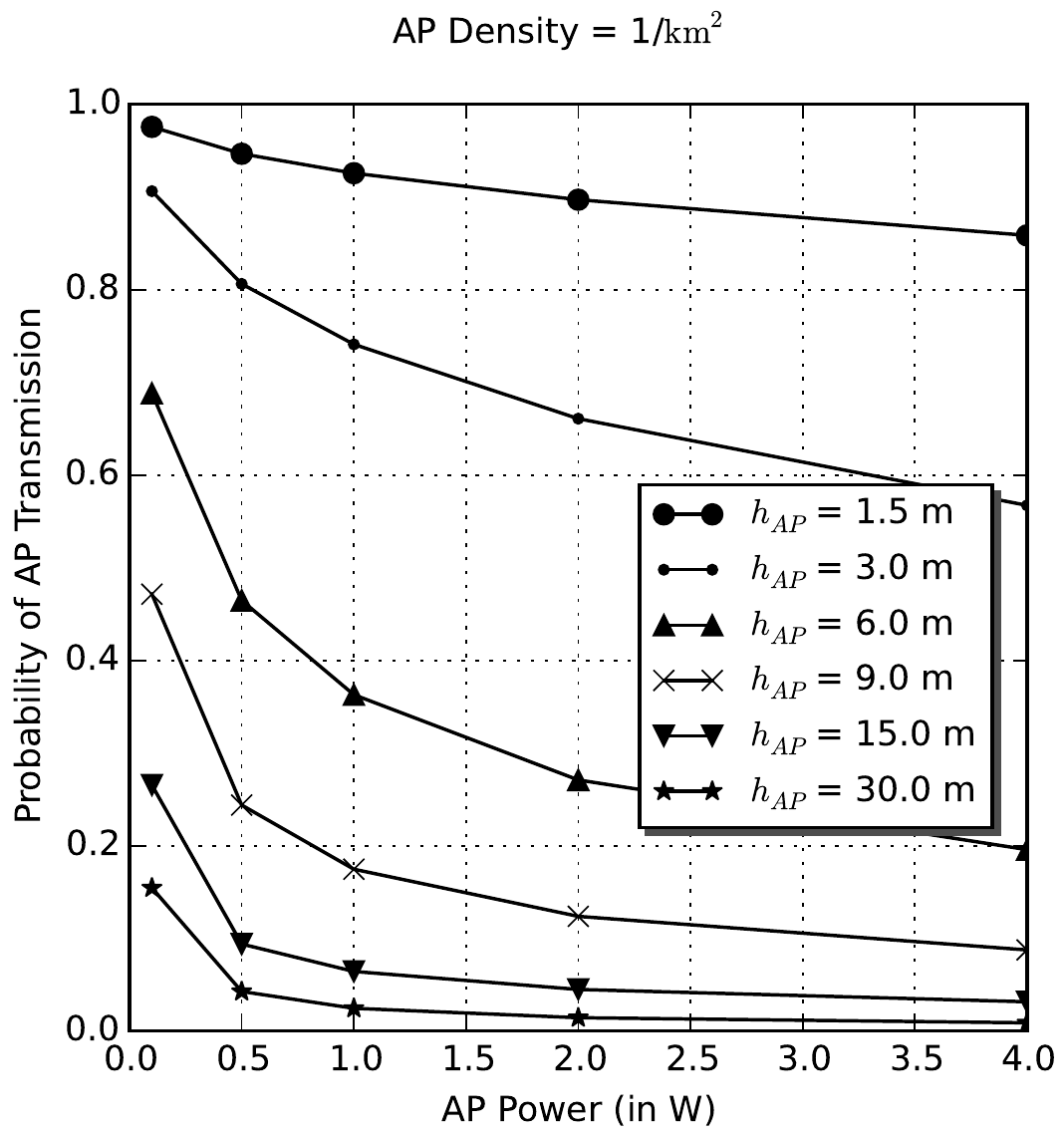}} \hfill
  \subfloat[High Density]{\label{pt_10}\includegraphics[width=5.9cm,height=6cm]{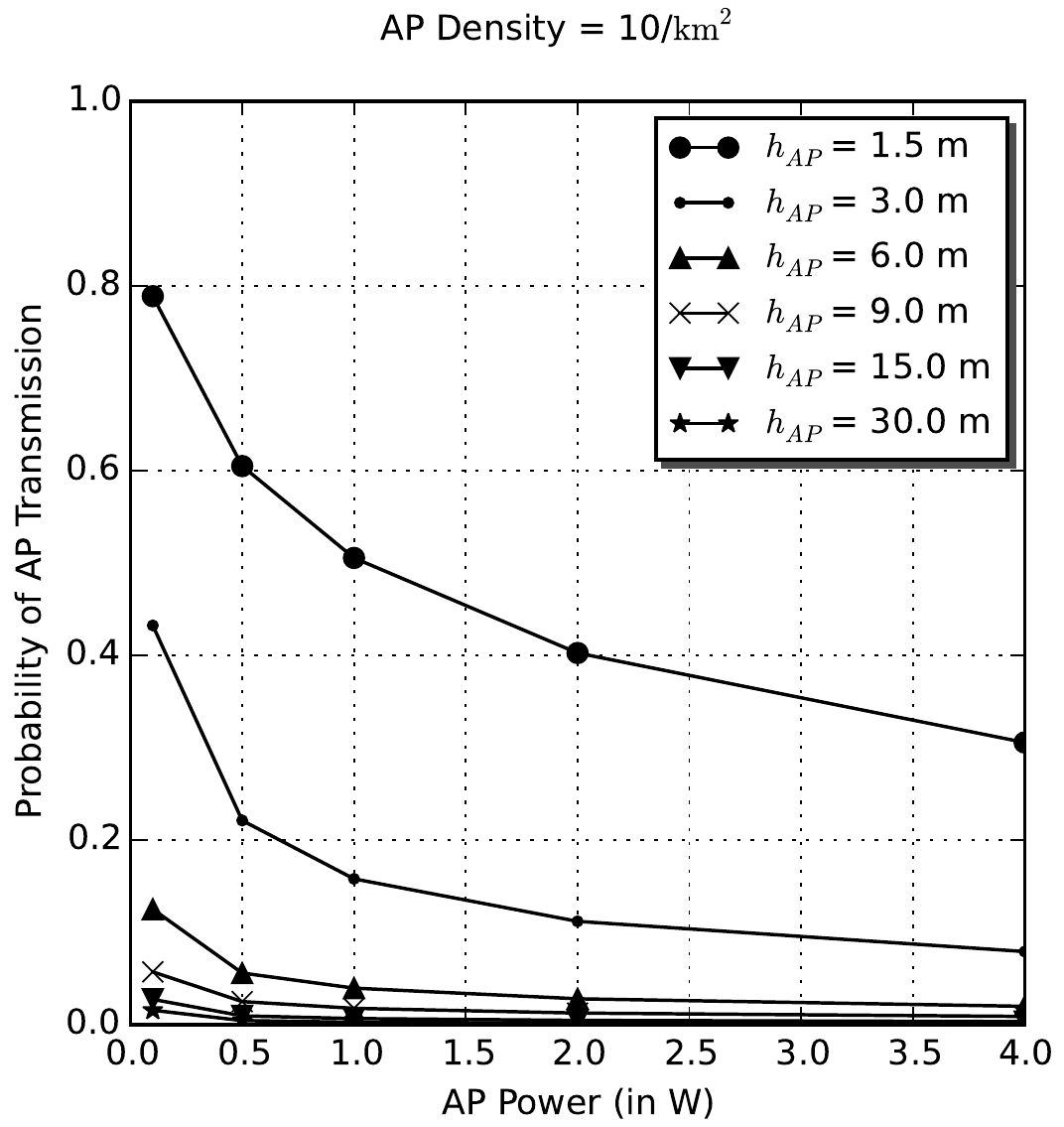}} \\ 
  \caption{Probability of transmission of an AP under different transmission powers $P_{AP}$, heights $h_{AP}$, and densities.}
  \label{PT}
\end{figure*} 
\begin{figure*}[t]
	\centering
	\subfloat[Probability of uplink viability vs. AP-client distance $\mathnormal{r}$ for various AP heights.]{\label{up_viability}\includegraphics[scale = 0.407]{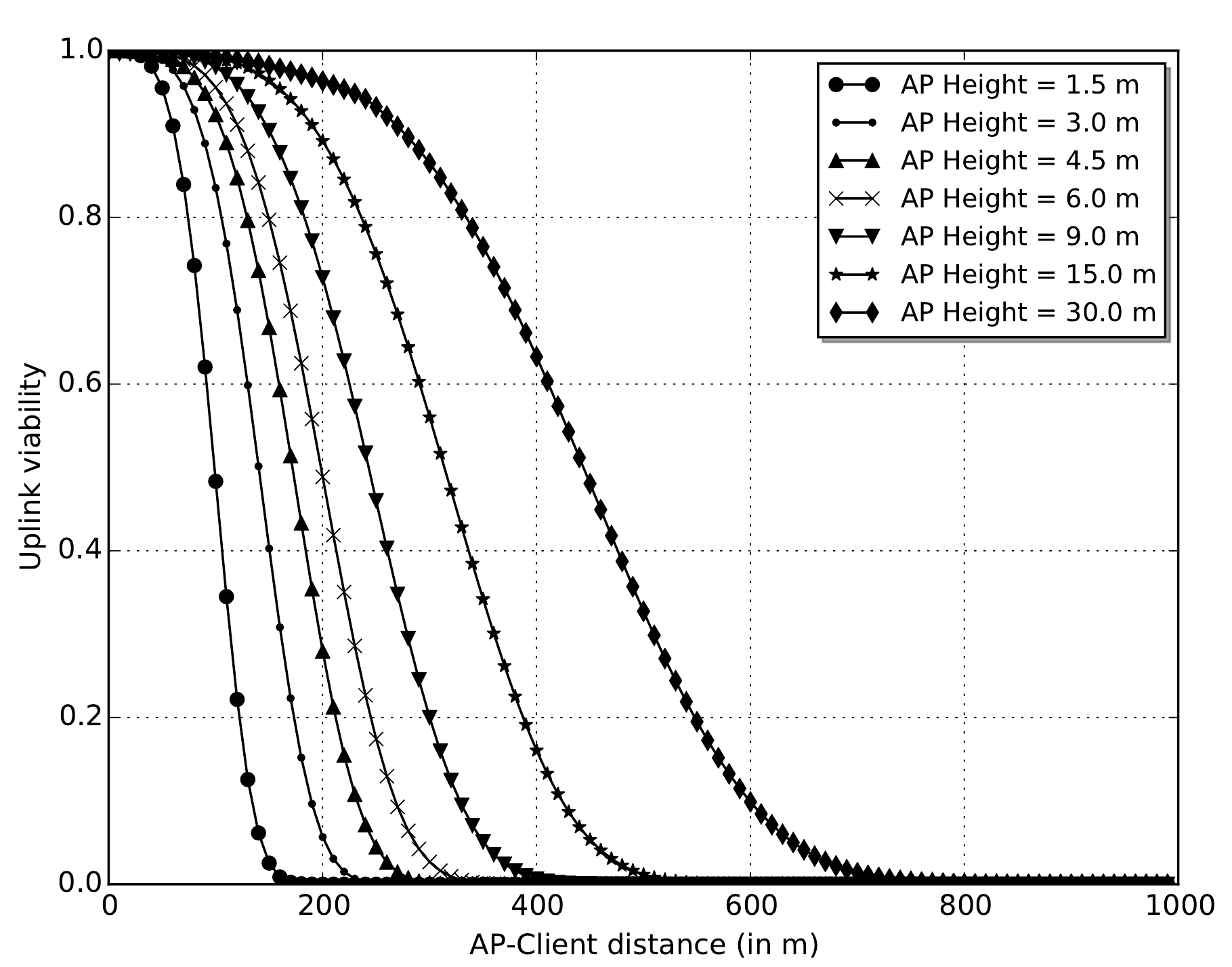}} \hfill
	\subfloat[Probability that client does not find an AP to associate with; note that client always transmits at 0.1 W.]{\label{starvation}\includegraphics[scale = 0.4]{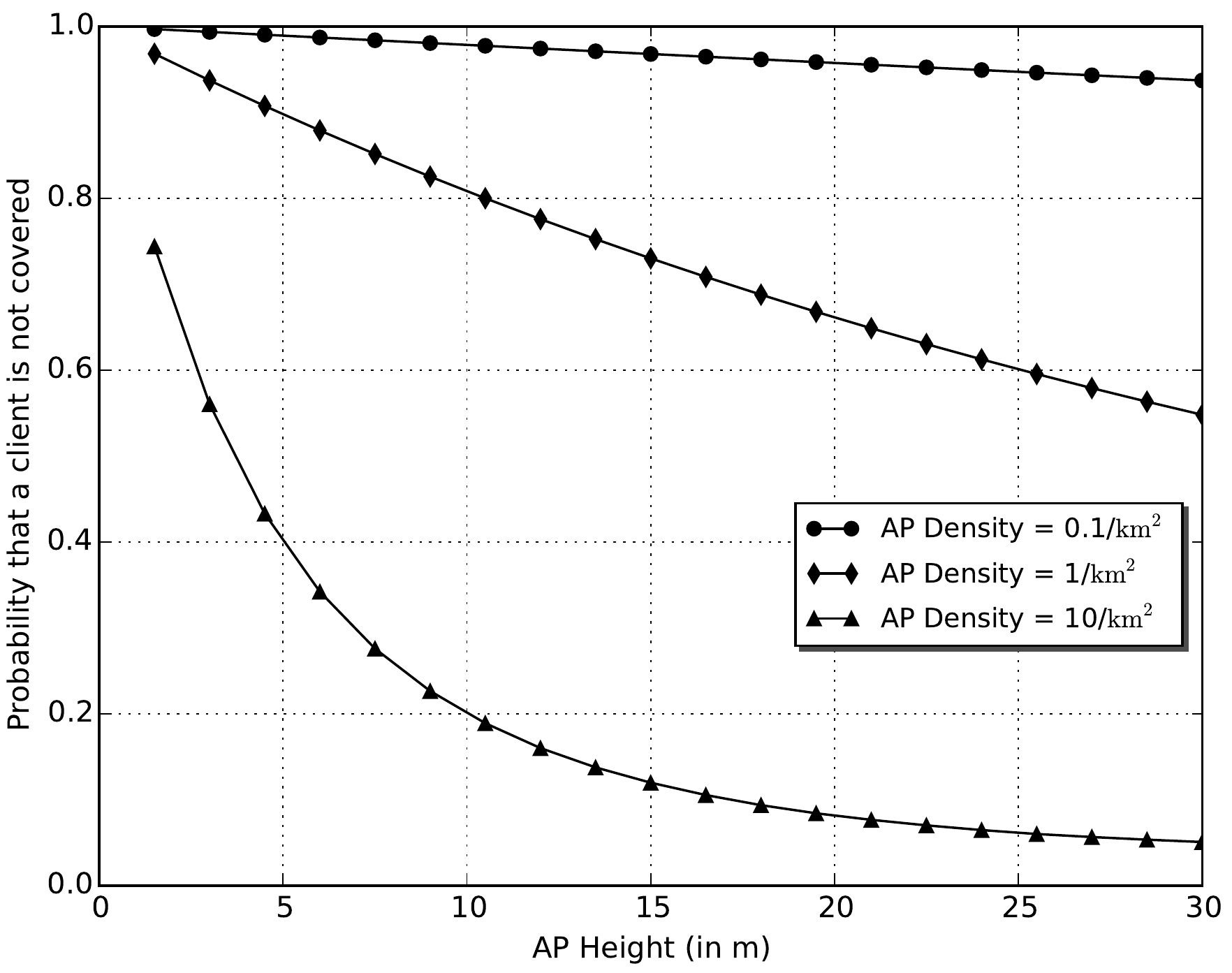} }
	\caption{Effect of AP density and height on uplink viability and geographic client coverage.}
	\label{p_uplink}
\end{figure*} 

\subsubsection{Transmission Probability}

This section examines the impact of AP transmit power ($P_{AP}$) and height ($h_{AP}$) on AP transmission probabilities at various deployment densities. Fig. \ref{PT} plots average $p_{T}$ as a function of $h_{AP}$ for three different densities. The average is computed over all AP-client distances $\mathnormal{r}$ and is given by  
\begin{equation} 
\bar{p}_T = \mathbb{E}_r[\mathnormal{p_T(r)}] = \int_{0}^{\infty} \mathnormal{p_T(r)} f_R(r|I_u=1) d\mathnormal{r}.
\end{equation} 

Fig.~\ref{PT} illustrates the interplay between  $P_{AP}$ and $h_{AP}$ in deciding how often an AP transmits and as expected, transmission probabilities decrease with increasing density. Note that at densities $\leq 1$ AP/km$^2$ and low antenna heights (1.5 to 3 m), transmit power does not play a significant role in determining the transmission probability. Further, it is seen that at higher transmission powers, $\bar{p}_T$ drops significantly as $h_{AP}$ increases. For instance, at an AP deployment density of 1 AP/$\mathrm{km^2}$, when $P_{AP} = 4\,\mathrm{W}$, an AP transmits with a probability of $\approx 0.85$ when operated at a height of 1.5 m but this probability drops below 0.1 when operated at a height of 15 m. In fact, at high densities ($\geq 10$ APs/km$^2$), operating at any height above 3 m does not seem optimal. On the other hand, given a target $\bar{p}_T$ and deployment density, and multiple $(P_{AP}, h_{AP})$ pairs that meet the target $\bar{p}_T$, choosing the pair with the highest AP height is advisable, as increasing AP height benefits both uplink and downlink, while increasing $P_{AP}$ only aids downlink, leading to greater asymmetry. Thus, when designing Super Wi-Fi networks, careful consideration must be given to the choice of $P_{AP}$ and $h_{AP}$, with particular attention paid to uplink-downlink asymmetry.

 

\subsubsection{Coverage Analysis} 
Unlike existing work on characterizing coverage by computing the probability that SINR exceeds a given threshold, this work defines coverage via uplink viability and determines a client to be in coverage if the uplink packets are received above the CCA threshold. When defined in this manner, coverage becomes independent of downlink transmit power. This alternate definition is particularly appropriate for Super Wi-Fi networks where clients are restricted to transmit at 0.1 W, but APs are allowed to transmit up to 4 W. Fig.~\ref{up_viability} plots uplink viability $p_U$ as a function of AP-client distance, assuming clients to transmit at 0.1 W. It can be observed that even at an AP height of 30 m, uplink is no longer viable beyond 700 m. In fact, for users who are 400 to 700 m away, uplink is viable less than $60\%$ of times---further restricting the range of an AP if reliable transmission is desired. This coverage limitation leads to large coverage gaps in a sparse deployment, as seen in Fig.~\ref{starvation}. Assuming clients to be uniformly distributed, Fig.~\ref{starvation} shows that at deployment densities of 1 AP/km$^2$, over 50\% of clients cannot associate with any AP. Thus, Super Wi-Fi networks with deployment densities less than 1 AP/km$^2$ are only able to provide localized coverage while densities greater than 10 APs/km$^2$ are required to ensure pervasive coverage over a wide area. 

\begin{figure*}[t] 
	\centering 
	\subfloat[$P_{AP} = 4$ W, $h_{AP} = 30$ m]{\label{coverage_0_1}\includegraphics[scale = 0.4]{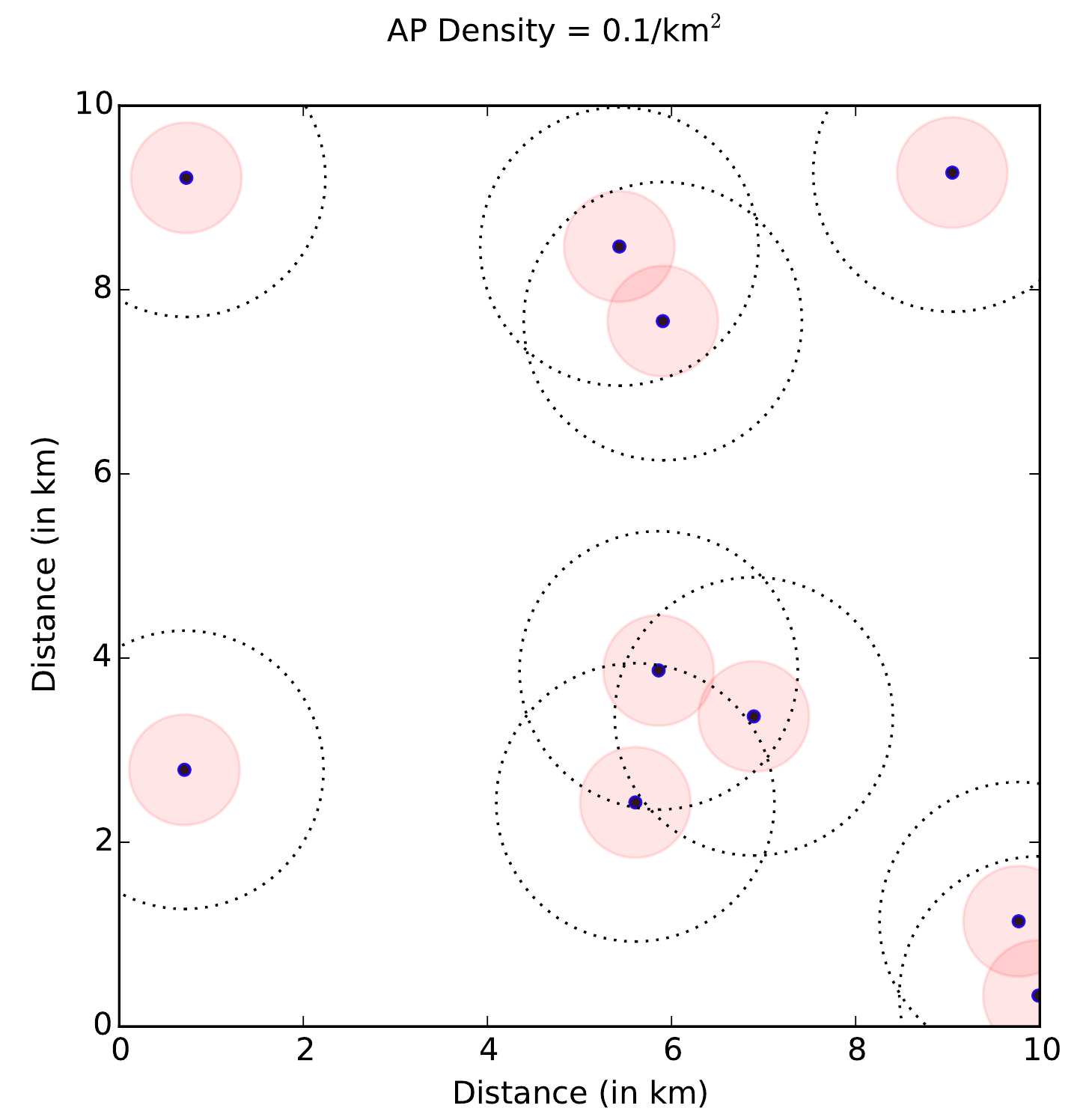}} \hspace{2cm}
	\subfloat[$P_{AP} = 1$ W, $h_{AP} = 10$ m]{\label{coverage_10}\includegraphics[scale = 0.4]{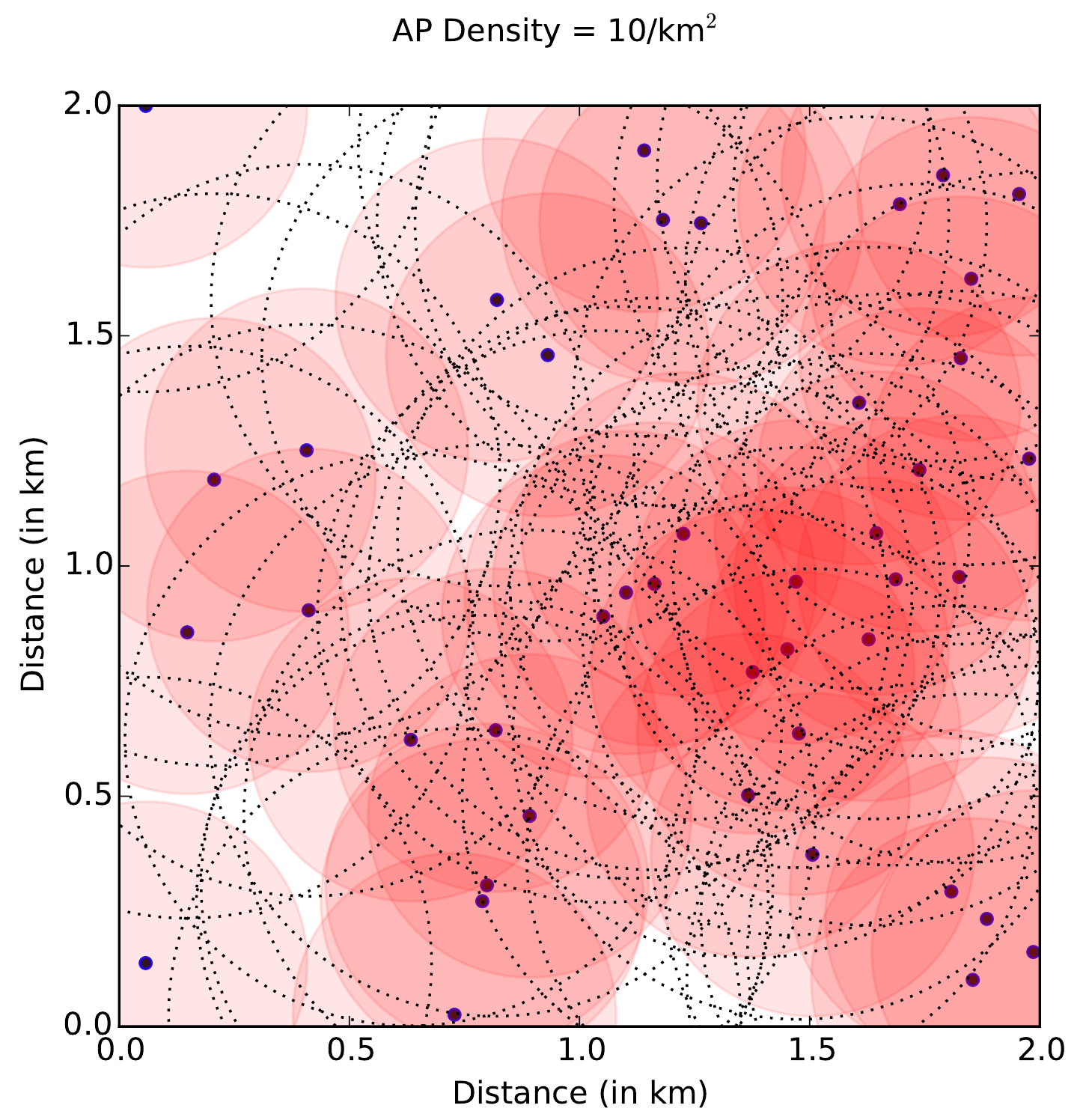}}
	\caption{AP downlink and uplink coverage for different deployment densities; dotted black circle - downlink coverage, red circle - uplink coverage.}
	\label{AP_coverage}
\end{figure*}
\begin{figure*}[t] 
	\centering 
	\subfloat[Low Density]{\label{ase_0_1}\includegraphics[width=5.9cm,height=6.2cm]{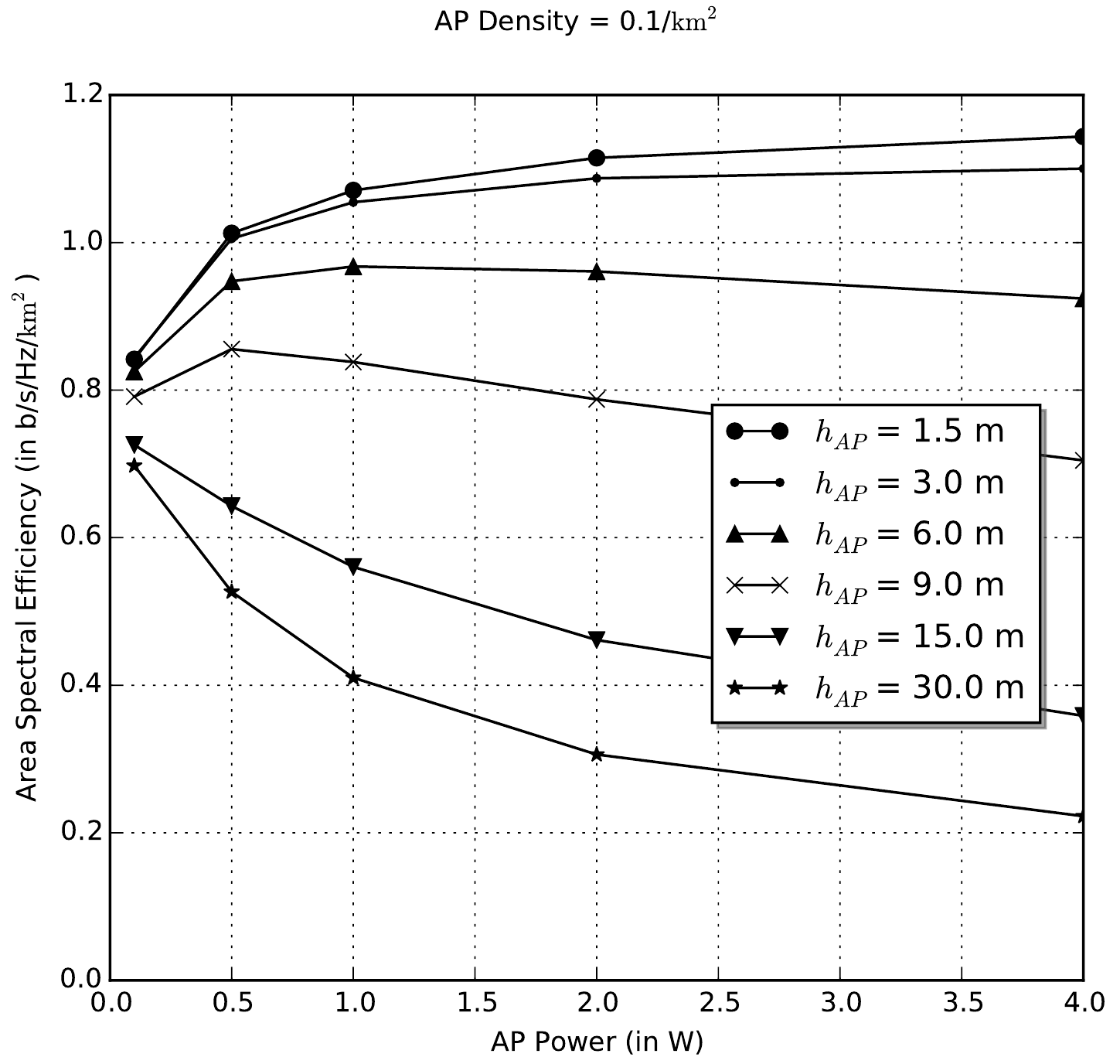}} \hfill
	\subfloat[Medium Density]{\label{ase_1}\includegraphics[width=5.9cm,height=6.2cm]{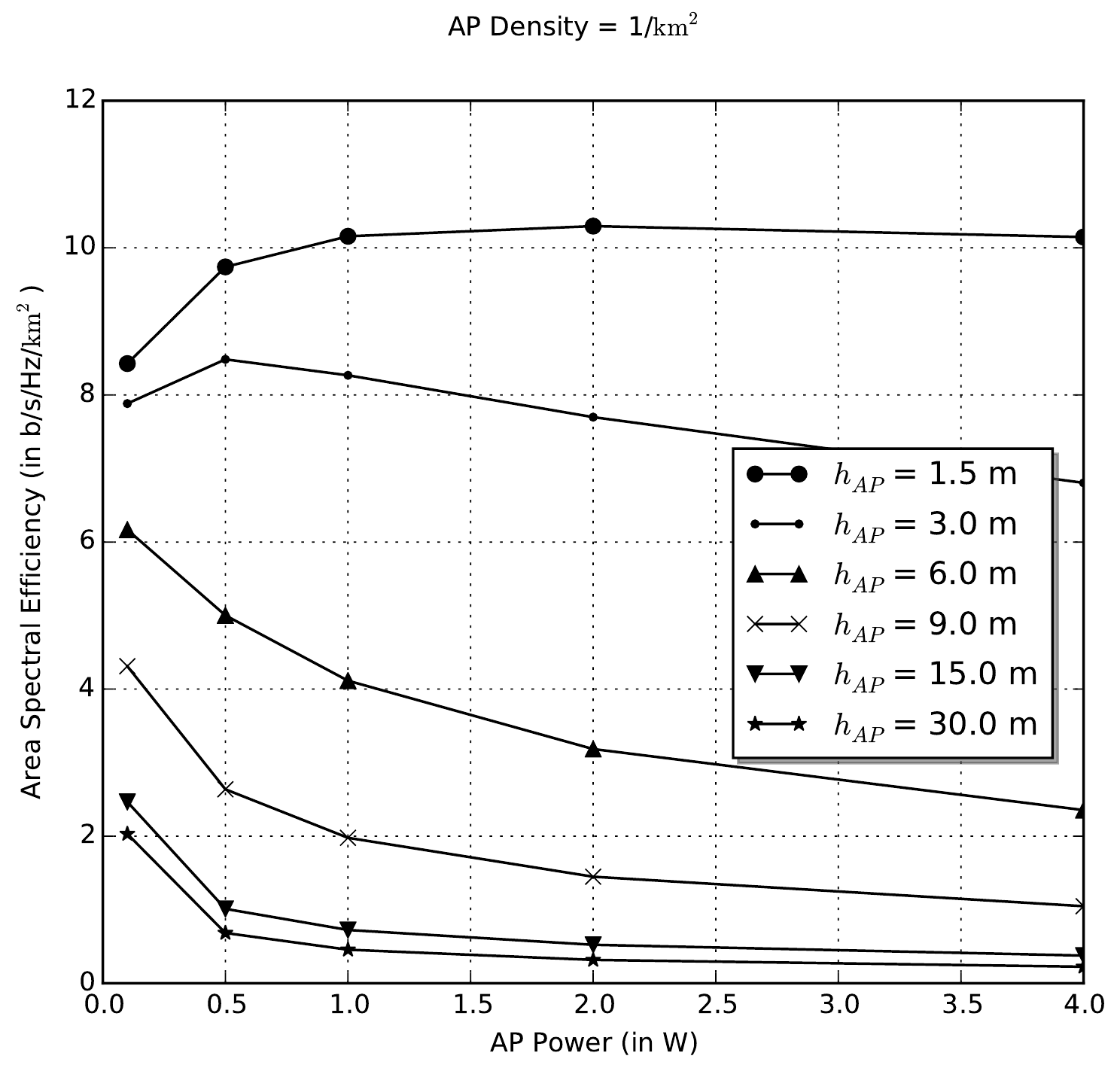}} \hfill
	\subfloat[High Density]{\label{ase_10}\includegraphics[width=5.9cm,height=6.2cm]{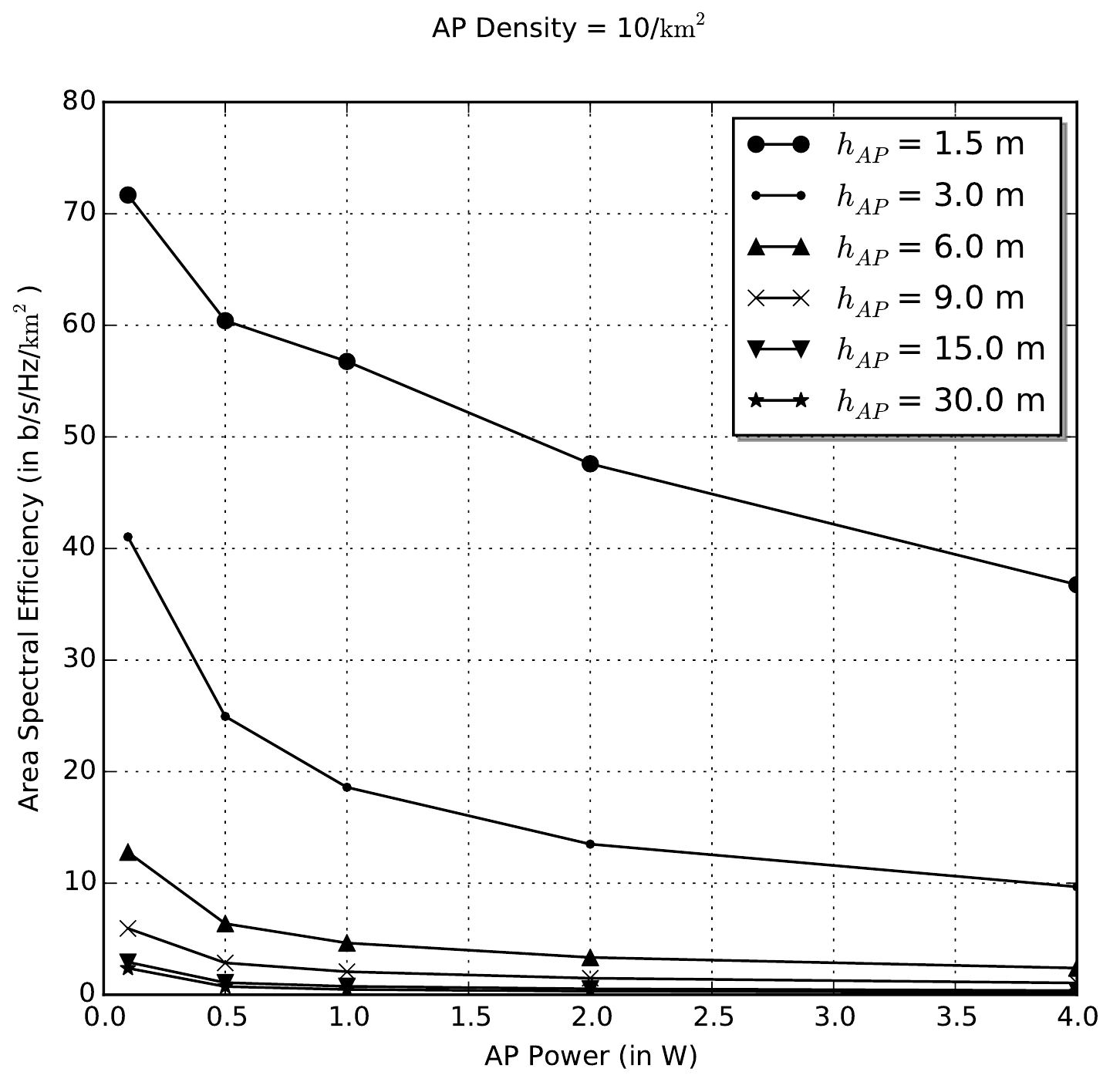}} 
	\caption{Area Spectral Efficiency (ASE) under different transmission powers $P_{AP}$, heights $h_{AP}$, and deployment densities.}
	\label{throughput}
\end{figure*} 
This observation is illustrated in Figs.~\ref{coverage_0_1} and \ref{coverage_10}. Fig.~\ref{coverage_0_1} shows coverage and downlink range (defined analogous to coverage range; determines channel contention radius) in a Super Wi-Fi network with $\lambda=0.1$ AP/km$^2$. APs are assumed to be at a height of 30 m and transmit at 4 W. It is clear that at such densities only localized coverage is possible. Further, the figure also illustrates why operating at $P_{AP} = 4$ W is not optimal even at such low AP deployment densities. High AP transmit powers lead to unnecessary enlargement of the contention radius, causing a drop in $\bar{p}_T$ and a potential decrease in network  performance (as seen in the next subsection).  Fig.~\ref{coverage_10} represents a Super Wi-Fi network with $\lambda=10$ APs/km$^2$, that is capable of providing pervasive coverage. Parameters $h_{AP}$ and $P_{AP}$ are set to 1~W and 10~m, respectively. At this height more than 80\% of the area is covered. Once again it is seen that higher transmit powers lead to a large contention radius that can be detrimental to network performance.

\subsubsection{Area Spectral Efficiency}


Throughput $\mathcal{T}$ of an AP is as defined in (\ref{tpt-2}). Area spectral efficiency is the product $\mathcal{T}\lambda$, and reflects the total number of bits transmitted over the wireless medium in a given area.
The following discussion is split into three cases, depending on the deployment density.

Fig.~\ref{ase_0_1}, plots performance of a very sparse deployment with a density of 0.1 AP/km$^2$. At such densities, only localized coverage is possible and for each curve in Fig.~\ref{ase_0_1}, the coverage characteristics remain the same. It is seen that for a fixed transmit power, ASE increases with decreasing height, owing to reduced coverage area, thereby serving only those clients who are at a very close proximity to the AP. On the other hand, if $h_{AP}$ is held fixed, then ASE increases with transmit power at lower heights, but decreases when $h_{AP}$ exceeds 9 to 10 m. This observation can be attributed to the fact that at lower AP heights, AP transmission probabilities are only a weak function of $P_{AP}$, and the gains in downlink SINR do not get negated by a drop in $\bar{p_T}$, as is the case for higher AP heights. In the case when maximum coverage is sought by setting $h_{AP}$ to 30 m, each AP delivers close to 40 Mbps over a 6 MHz TVWS channel when operating at a transmit power of 0.1 W. The key takeaway here is that when seeking to maximize localized coverage (by setting $h_{AP}>10$ m), increasing AP transmit power is unlikely to yield better performance due to the sharp drop in AP transmission probability. 

Fig.~\ref{ase_1} considers a medium deployment density of 1 AP/km$^2$.  Once again two different behaviors are seen depending on whether $h_{AP}$ exceeds 3 m or not. It is clear that $h_{AP}$ exceeding 10 m has a detrimental impact on network performance. Even at such densities, pervasive coverage is difficult to achieve. When maximum coverage is sought, each AP delivers close to 12 Mbps per TVWS channel, with a total of 120 Mbps/km$^2$. 

Fig.~\ref{ase_10} considers a dense deployment scenario with a density of 10 APs/km$^2$. In this case, it is possible to achieve pervasive coverage when $h_{AP}$ exceeds 15 m. Unlike the previous two cases, for a fixed $h_{AP}$, ASE decreases with increasing $P_{AP}$ suggesting that changes in $\bar{p}_T$ plays a more important role than changes in SINRs. Maximizing coverage is not as important as before, and setting 9 m $\leq h_{AP} \leq$ 15 m suffices to ensure that more than 80\% of the clients are under coverage. When $h_{AP} = 9$ m, an ASE of 240 Mbps/km$^2$ over one TVWS channel can be achieved. The striking similarity between Fig.~\ref{pt_10} and Fig.~\ref{ase_10} suggests that this network is interference limited where interference dominates over noise and an increase in transmit power leads to an equal amount of increase in signal and interference strength, leaving SINR unchanged. 

The key takeaways from this section are summarized in Table~\ref{summary} where $\lambda$ denotes the number of APs/km$^2$. Note that pervasive coverage is achieved only at high deployment density and when coverage per AP is prioritized. 
\begin{table}[t]
\caption{Suggested choice of AP operating parameters for different deployment densities}
\label{summary} 
\vspace{-0.3cm}
\begin{center}
\begin{tabular}{|c |c |c |}
\hline 
Priority $\rightarrow$ & Max. coverage/AP & Max. throughput/AP \\
\hline 
Sparse ($\lambda = 0.1$) & (0.1 W, 30 m) & (1 to 4 W, 1.5 to 3 m) \\ 
\hline 
Medium ($\lambda = 1$) & (0.1 W, 30 m) & (1 to 4 W, 1.5 m) \\ 
\hline 
High ($\lambda = 10$) & (0.1 W, 10 to 30 m) & (0.1 W, 1.5 m) \\ 
\hline 
\end{tabular}
\end{center}
\end{table} 

\subsection{Network Planning for a Suburban use-case} 
\begin{figure}[t]
\centering 
\includegraphics[scale = 0.4]{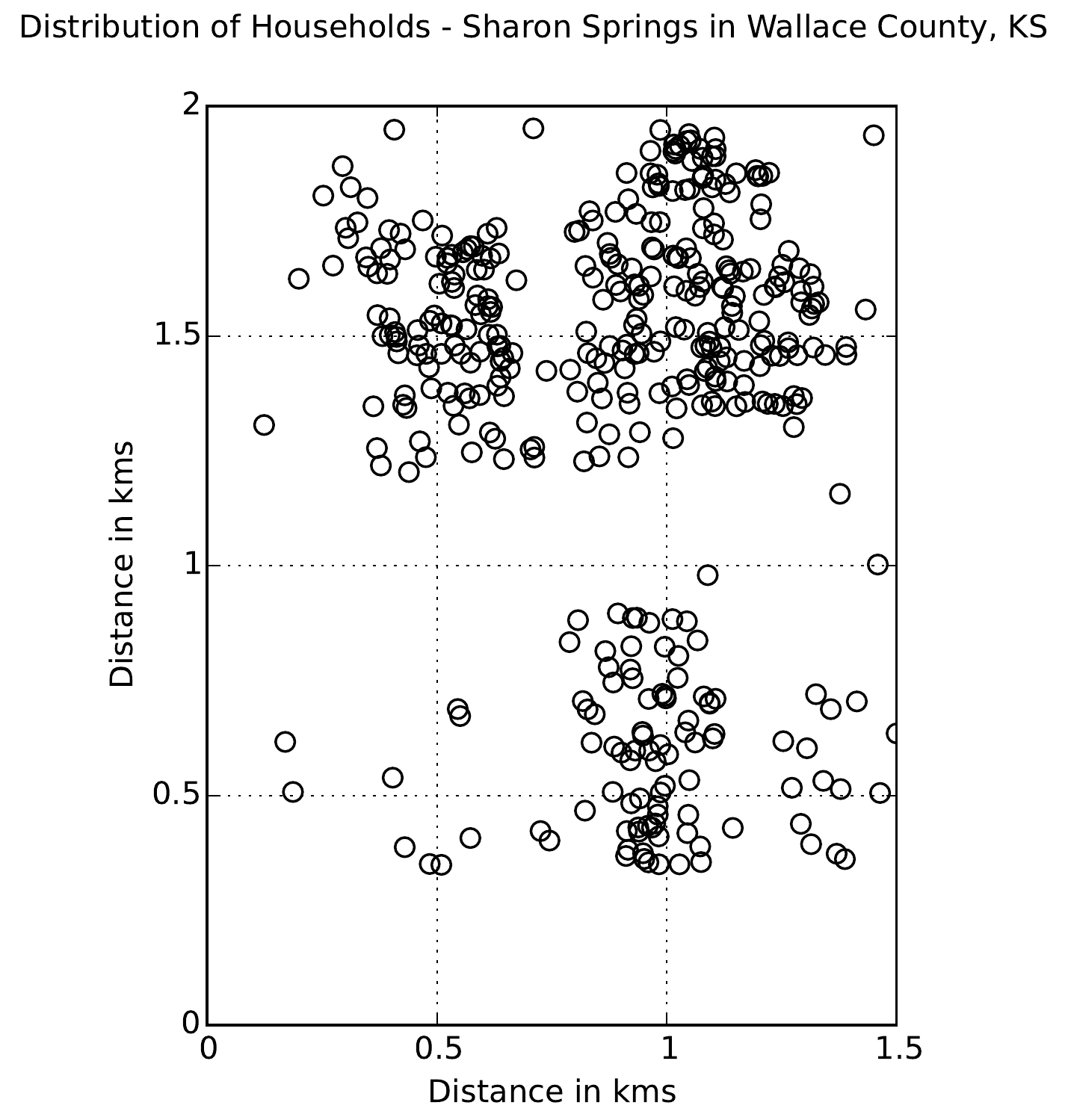}
\caption{Distribution of households in Sharon Springs, Wallace County, Kansas according to 2010 US Census. Each circle represents a house.} 
\label{wallace}
\end{figure}

This section explores a network plan for providing broadband connectivity to a suburban region using results from the previous discussion. The region of interest is Sharon Springs in Wallace County, Kansas, USA. 
Fig.~\ref{wallace} shows the distribution of houses in Sharon Springs obtained from the records of 2010 US Census. 
There are around 400 households spread out over an area of 3 km$^2$. Suppose each household is to be supported with a data rate of 10 Mbps, the required network throughput is $400\times10/3 = 1330\, \text{Mbps/km}^2$. In this area, 37 TVWS channels amounting to a total of 222 MHz are currently available (from Google spectrum database \cite{googlelink}).

Consider deploying a Super Wi-Fi network with a deployment density of 10 APs/km$^2$. With the objective of attaining a coverage probability of at least 75\% and with the ease of mounting antennas on street light poles in mind, AP height is chosen to be $h_{AP} = 6$ m. Using results from Fig.~\ref{ase_10}, AP transmit power $P_{AP}$ is set to 0.1 W to obtain an ASE of 12~bps/Hz/km$^2$, translating to a network throughput of 72~Mbps/km$^2$ per TVWS channel. Thus, it is possible to meet the demands of this suburban region using at most 18 of the 37 available channels. Further network efficiency and better coverage can be achieved using a more careful AP deployment leading to a more economical use of available bandwidth. 


These computations suggest that such a network can be a reasonable access alternative to satellite-based internet service which tends to be the dominant means of connectivity in such rural areas. Backhaul services for the deployed APs can also be provided using TVWS channels as investigated in \cite{cyrus,sudhir}. 

\section{CONCLUSION} 
\label{conclusion}
This paper uses a stochastic geometry analysis to study the feasibility of utilizing TVWS channels to provide broadband connectivity in rural and under-served regions using a Wi-Fi-like network. Regulations on transmit power and antenna height for both APs and clients present situations in which the downlink may be viable but the uplink from client to AP is not. The performance of such a network operating in TVWS channels is analyzed using stochastic geometry while explicitly factoring in uplink viability. Such an analysis is used to characterize AP transmission probabilities, coverage, and area spectral efficiency. These results show that operating APs at high transmit powers and heights is not always beneficial to the performance of the network, even at low AP deployment densities. It is however seen that APs deployed at higher heights significantly improve uplink viability. This exemplifies the rate-coverage trade-off in these networks. The choice of operating parameters for such a network will depend on the desired balance between coverage and rate. 

\section*{ACKNOWLEDGMENT}
This work is supported in part by a grant from the U.S. Office of Naval Research (ONR) under grant number N00014-15-1-2168. 

\appendices
\section{Derivation of probability of concurrent AP transmissions}
Let AP $\vec{x}$ represent an AP that is at a distance $d$ from AP $\vec{0}$. Then, $\mathnormal{q(d)}$ represents the probability that AP $\vec{0}$ and AP $\vec{x}$ transmit at the same time, and can be  written as
\begin{align}
\mathnormal{q(d)} =& \mathbb{P}_{\Phi_A}^{\vec{0},\vec{x}} \left\{\vec{x} \in \Phi_T | \vec{0} \in \Phi_T \right \}  \nonumber \\
=& \frac{\mathbb{P}_{\Phi_A}^{\vec{0},\vec{x}} \left\{\vec{x} \in \Phi_T , \vec{0} \in \Phi_T \right \}}{\mathbb{P}_{\Phi_A}^{\vec{0},\vec{x}} \left\{\vec{0} \in \Phi_T \right \}}.
\label{q_eq}
\end{align}
To compute the numerator of (\ref{q_eq}), let $m_\vec{0}$ and $m_\vec{x}$ as the marks chosen by AP $\vec{0}$ and AP $\vec{x}$ respectively and assume $m_\vec{0} < m_\vec{x}$, without loss of generality. Denote $\vec{z}$ as a potential interferer. To compute the joint probability that both AP $\vec{0}$ and AP $\vec{x}$ concurrently transmit, two `classes' of APs need to be considered---those with mark $m < m_\vec{0}$, distributed as a PPP of intensity $\lambda m_\vec{0}$, which prevent both AP $\vec{0}$ and AP $\vec{x}$ from transmitting, and those with mark $m_\vec{0} < m < m_\vec{x}$, distributed as a PPP of intensity $\lambda (m_\vec{x} - m_\vec{0})$, which prevent only AP $\vec{x}$ from transmitting. Using these observations, the numerator of (\ref{q_eq}) can be computed as
\begin{align}
&\mathbb{P}_{\Phi_A}^{\vec{0},\vec{x}} \left\{\vec{x} \in \Phi_T , \vec{0} \in \Phi_T\right\} \nonumber \\
&\hspace{-1.2mm}=2(1-e^\frac{-\mu\sigma}{P_{AP}\mathnormal{PL(x)}})\int_{0}^{1}\left [\int_{\mathnormal{m_\vec{0}}}^{1} e^{-\lambda(\mathnormal{m_\vec{x}-m_\vec{0}})  \int_{\mathbb{R}^2} \mathnormal{S_\vec{x}}(\vec{z}) d\vec{z}}d\mathnormal{m}_\vec{x} \right ] \nonumber \\
&\phantom{=} \hspace{3cm} \times e^{-\lambda\mathnormal{m_\vec{0}}\int_{\mathbb{R}^2} \mathnormal{S_{\vec{0} \, or \, \vec{x}}}(\vec{z}) d\vec{z}} d\mathnormal{m_\vec{0}} 
\end{align}
%
where $1-e^\frac{-\mu\sigma}{P_{AP}\mathnormal{\rho(\vec{x},\vec{0})}}$ is the probability that AP $\vec{x}$ is not in the neighborhood of AP $\vec{0}$. The factor of two accounts for the case when $m_\vec{x} < m_\vec{0}$. $\mathnormal{S}_{\vec{x}}(\vec{z})$ is the probability that AP $\vec{x}$ senses the transmission of AP $\vec{z}$ and $\mathnormal{S_{\vec{0} \, or \, \vec{x}}}(\vec{z}) = 1 - (1-S_\vec{x}(\vec{z}))(1-S_\vec{0}(\vec{z}))$ is the probability that the interfering AP $\vec{z}$ is sensed by at least one of AP $\vec{0}$ or AP $\vec{x}$. 
\\
In a similar manner, the denominator of (\ref{q_eq}) can be computed as 
\begin{equation} 
\begin{split}
\mathbb{P}_{\Phi_A}^{\vec{0},\vec{x}} \left\{\vec{0} \in \Phi_T \right \} = \int_{0}^{1}\int_{0}^{\mathnormal{m}_\vec{x}} e^{-\lambda \mathnormal{m}_\vec{0}\int_{\mathbb{R}^2}\mathnormal{S}_\vec{0}(\vec{z}) d\vec{z}} d\mathnormal{m}_\vec{0}\, + \\
\int_{\mathnormal{m}_\vec{x}}^{1} (1-e^{\frac{-\mu\sigma}{P_{AP}\mathnormal{PL(x)}}})e^{-\lambda\mathnormal{m}_\vec{0}\int_{\mathbb{R}^2}\mathnormal{S}_\vec{0}(\vec{z}) d\vec{z}} d\mathnormal{m}_\vec{0}   \, d\mathnormal{m}_\vec{x}.
\end{split}
\end{equation}
\\
In the above expression, the first term considers the case when $m_\vec{0} < m_\vec{x}$, where AP $\vec{0}$ can transmit regardless of whether AP $\vec{x}$ is transmitting or not, while the second terms considers the case when $m_\vec{0} > m_\vec{x}$, in which case, AP $\vec{0}$ can transmit only if the AP $\vec{0}$ cannot sense AP $\vec{x}$'s transmissions.

\bibliographystyle{IEEEtran}
\bibliography{IEEEabrv,bibliography_paper}
\end{document}